\documentclass[10pt]{article}
\usepackage{graphicx}

\usepackage{epsfig}
\usepackage{latexsym}
\usepackage{amsmath}

\begin{document}

\title{\Large \bf Spatial interactions in agent-based modeling  }

\author{ \large \bf Marcel Ausloos$^{1,2,3,}$\footnote {marcel.ausloos@ulg.ac.be,} ,  Herbert Dawid$^{4,}$\footnote {hdawid@wiwi.uni-bielefeld.de,} ,    Ugo Merlone$^{5,}$\footnote {corresponding author: ugo.merlone@unito.it} \\
\\$^1$ R\'es. Beauvallon, rue de la Belle Jardini\`ere, 483/0021, \\
B-4031, Li\`ege Angleur, Euroland 
\\ $^2$ GRAPES, rue de la Belle Jardiniere, B-4031 Liege, \\Federation Wallonie-Bruxelles, Belgium
\\ $^3$ e-Humanities Group, KNAW,\\Joan Muyskenweg 25, 1096 CJ Amsterdam, The Netherlands \\
\\$^4$ Department of Business Administration and Economics \\and \\Center for Mathematical Economics,  \\Bielefeld University, 
Universit\"{a}tsstra\ss e 25\\
D-33615 Bielefeld, Germany \\
\\ $^5$ Department of Psychology, Universit\`{a} di Torino, \\via Verdi 10, Torino I-10124, Italy \\  }

\maketitle

\abstract {Agent Based Modeling (ABM) has become a widespread approach to
model complex interactions. In this chapter after briefly summarizing some features
of ABM the different approaches in modeling spatial interactions are discussed. \\
It is stressed that agents can interact either indirectly through a shared environment and/or directly with each other. In such an approach,  higher-order variables such as commodity prices, population dynamics or even institutions, are not exogenously specified but instead are seen as the results of interactions. It is highlighted in the chapter that the understanding of patterns emerging from such spatial interaction between agents is a key problem as much as their description through analytical or simulation means. \\
The chapter reviews different approaches for modeling agents' behavior, taking into account either explicit spatial (lattice based) structures or networks.  Some emphasis is placed on  recent  ABM  as applied to the description of the dynamics of the geographical distribution of economic activities, - out of equilibrium.  The Eurace@Unibi Model, an agent-based macroeconomic model with spatial structure,  is used to illustrate the potential of such an approach for spatial policy analysis.}

\textbf{Abstract} Agent Based Modeling (ABM) has become a widespread approach to
model complex interactions. In this chapter after briefly summarizing some features
of ABM the different approaches in modeling spatial interactions are discussed. 

It is stressed that agents can interact either indirectly through a shared environment and/or directly with each other. In such an approach,  higher-order variables such as commodity prices, population dynamics or even institutions, are not exogenously specified but instead are seen as the results of interactions. It is highlighted in the chapter that the understanding of patterns emerging from such spatial interaction between agents is a key problem as much as their description through analytical or simulation means. 

The chapter reviews different approaches for modeling agents' behavior, taking into account either explicit spatial (lattice based) structures or networks.  Some emphasis is placed on  recent  ABM  as applied to the description of the dynamics of the geographical distribution of economic activities, - out of equilibrium.  The Eurace@Unibi Model, an agent-based macroeconomic model with spatial structure,  is used to illustrate the potential of such an approach for spatial policy analysis.

\section{Agent-Based Modeling}
\label{sec:1}

In recent years there has been a lot of excitement about the potential of
agent-based modeling (ABM). We briefly summarize the ABM approach and mention some of its applications.

In agent-based modeling, a system is modeled as a
collection of autonomous decision-making entities called agents \cite{Bonabeau2002}.
Each agent individually assesses its situation and makes decisions on
the basis of a set of rules. 

When it comes to actual models, different approaches are proposed. For example, Axelrod proposed the KISS principle \cite[p.5]{Axelrod1997}. This principle comes  from the old army slogan, ``Keep it simple, stupid'' and is obviously related to the Occam's razor \cite{lazar2010ockham}. When dealing with complex systems this principle  is vital as, when surprising results are discovered, it is quite helpful to be confident that everything  can be understood in the model that produced the surprises \cite{Axelrod2000}.   Yet, other authors have opposite views and advocate more descriptive approaches, see  \cite{EdmondsMoss2005}. 

Given the level of details which can be used to model agents, this discussion is reflected also in how agents' behaviors are modeled. Several authors used data gathered in experiments, see for example  \cite{DalFornoMerlone2004,BoeroEtAl2010}. Recently, other approaches advocated the use of grounded theory to model agents' behavior, which enables the use of both quantitative and qualitative data, see \cite{AndrewsEtAl2005,DalFornoMerlone2006k,DalFornoMerloneWSC2012,daha11}.

\begin{figure}[b]
\includegraphics[scale=.45]{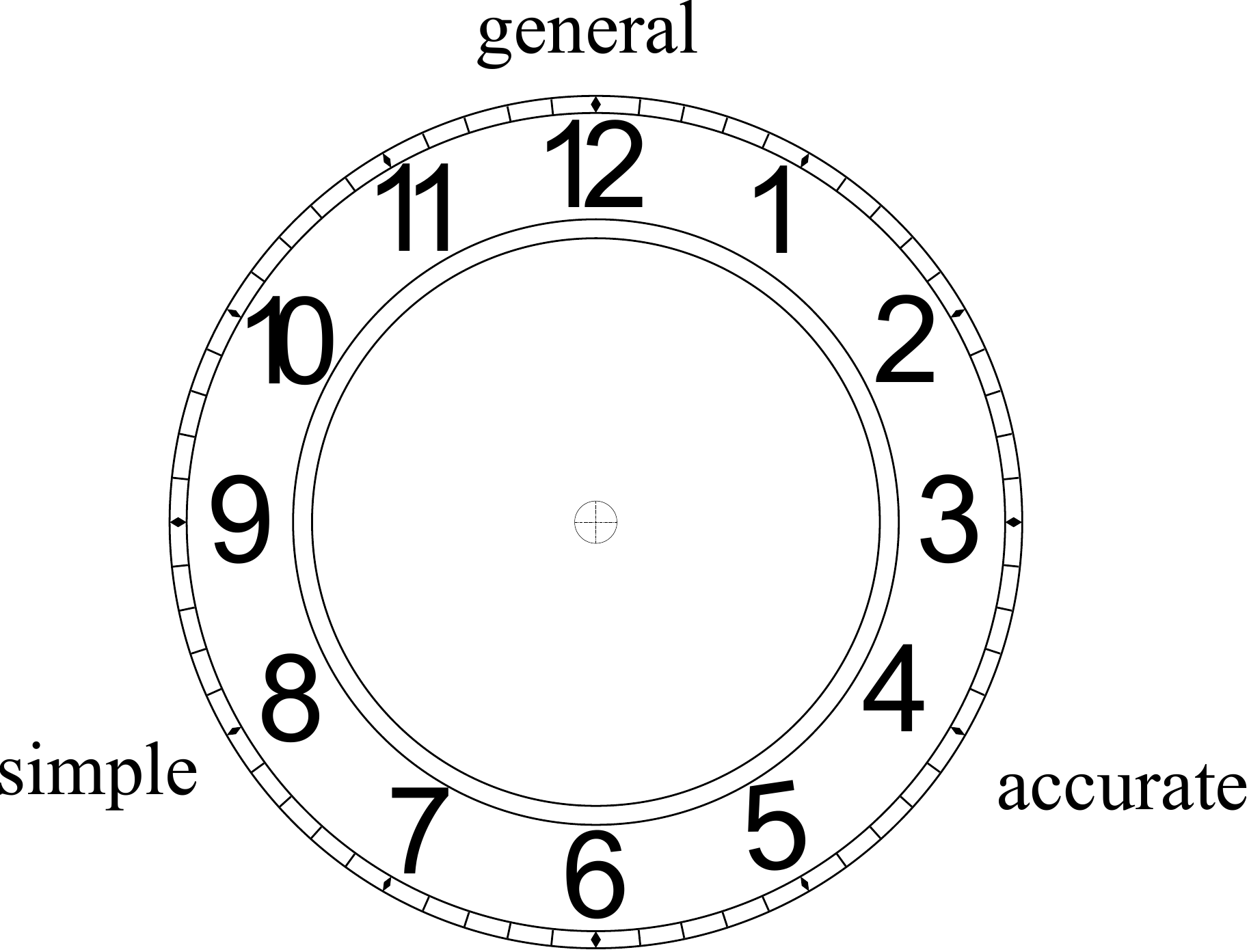}
%
%
\caption{Thorngate's ``postulate of commensurate complexity'' \cite{Thorngate1976} as represented by Weick's   \cite{Weick1979} clock metaphor  }
\label{fig:1}       
\end{figure}

Nevertheless to put things in the right perspective it should be kept in mind Thorngate's
 ``postulate of commensurate complexity'', i.e.  it is
impossible for a theory of social behavior to be simultaneously
general, accurate, and simple; as a result organizational theories
inevitably have  tradeoffs in their  development \cite{Thorngate1976}. To illustrate this postulate 
Karl Weick 
 \cite{Weick1979}   proposed the clock metaphor which is illustrated in Figure \ref{fig:1}.

This metaphor uses a clockface with \emph{general}
at 12:00, \emph{accurate} at 4:00, and \emph{simple} at 8:00 and shows that  an explanation satisfying any two characteristics is least able to satisfy the third characteristic.

ABM has been used for theory building in social psychology \cite{SmithConrey2007},  to model social processes as interactions \cite{MacyWiller2002}; according to \cite{Bonabeau2002} this approach is appropriate when considering  emergent phenomena in the social, political, and economic sciences. In particular in a business context, situations of interest where emergent phenomena may arise can be  flow simulation, organizational simulation, market simulation, and diffusion simulation. 

In agent-based modeling (ABM), a system is modeled as a collection of autonomous decision-making entities called agents which  interact
both with each other and with their environment.
The behavior of the whole system
is the result of the aggregated individual behavior of
each agent. 
Agents can interact either indirectly through a shared environment and/or directly with each other

This way, higher-order variables such as commodity prices, population dynamics or even  institutions are
not specified  but, instead, are the result of interaction, i.e., emergent outcomes.

\section{ABM Compared to Other Approaches}
\label{sec:ABMComp}
As observed in \cite{OSullivanEtAL2012}  ABMs are a relatively late arrival in fields where there is considerable previous experience with styles of model that adopt a more aggregated approach.

Also, in most fields the aggregated approach used so far  continues to be quite common, therefore it would be useful to compare and assess the ABM potentialities and critical aspects, when contrasted to other approaches. 
In \cite{GilbertTroitzsch2005} for example, ABM are compared to other social science simulation techniques in terms of communication between agents complexity of agents, number of agents and number of levels of interaction.

 Related to the study of socio-economic problems implying space and time, one can find in the condensed matter literature several models of evolution, in particular modeling crystal growth. In such cases, reaction rates and different degrees of freedom coupled to their corresponding external field  serve by analogy to describe the evolution of a society.  Among such  studies, several papers  can be mentioned, as in \cite{ausloos1996growth,vandewalle1994competition,vandewalle1995evolution}
 
In such  evolutionary economics models,  economic agents randomly search for new technological design by trial-and-error and run the risk of ending up in sub-optimal solutions due to interdependencies between
the elements in a complex system. As argued by Frenken  \cite{frenken1999interdependencies,frenken2001modelling},  these models of random
search are legitimate for reasons of modeling simplicity, but remain limited in scope   as these
models ignore the fact that agents can apply heuristics.

  It has been searched within agent-based model frameworks to provide an analogy between share price instabilities and 
fluctuations or  
instabilities in electrical circuits   which 
fluctuate in the vicinity of an unstable point \cite{glansdorff1971structure}, i.e.,  to provide the
price as a thermodynamic-like variable. Recent observations have indicated that the traditional equilibrium market hypothesis (EMH; also known as  Efficient Market Hypothesis) is unrealistic. It has been shown  in  \cite{ausloos2000gas}  that  the EMH  is the analog of a Boltzmann equation in physics, - thus having some bad properties of mean-field approximations,  e.g. 
  a Gaussian distribution of price  fluctuations, rather than the empirically found ``fat tails''  \cite{mantegna1999introduction}.  A better kinetic theory for prices can be simply derived and solved,  within a Chapman-Enskog-like formalism, considering in a  first approach  that market agents have all identical relaxation times.  In closing the set of equations, (i) an equation of state with a pressure and (ii) the equilibrium (isothermal) equation for the price (taken as the order parameter) of a stock as a function of the volume of money available are
obtained.
 
 The  Boltzmann kinetic equation  idea has been extended  in \cite{gligor2002kinetic} to describe  an idealized system composed by many individuals (workers, officers, business men, etc.), each of them getting a certain income and spending money for their needs. To
each individual a certain time variable amount of money was associated, -  this  defining him/her phase
space coordinates. In this  approximation,  the exponential distribution of money in a closed economy was explicitly
found. The extension of this result, including states near the equilibrium,  has given the possibility to take
into account the regular increase of the total amount of money, according to modern economic
theories.

 Other approaches consider with more detail the behavioral aspects of interaction. For example in  \cite{terna_jes} learning and adaptation are considered and interesting emerging characteristics of the agents can be observed. Also in \cite{terna2009epidemic} ABMs are used to understand and modify firms' behavior and to suggest possible solutions to  real life applications. 

\section{Spatial Interactions}
\label{sec:SpatInt}
Spatial interaction is an important topic, see for instance \cite{power,Stanilov2012}. In fact the interactions agents have depend on where agents are situated. Modeling the environment in which interaction takes place assume therefore an important role in modeling important phenomena. In the following we provide a rough classification of the spatial structures where interaction takes place providing examples which how flexible ABM is to deal with different spatial structures.

\subsection{No Explicit Spatial Structure}
\label{subsec:NoStr}
In several models there is no explicit spatial structure in which the interaction among agents takes place. Rather, agents are considered being part of a unique population. This kind of approach can be considered  to be a particular case of the spatial interactions that will be considered in Subsections  \ref{subsec:GeomInt} and \ref{subsec:NetwInt}. Nevertheless, even when no explicit spatial structure is considered interactions among heterogeneous agents have been analyzed in agent-based modeling. One important case is given by the the $N$-person prisoner's dilemma game which, according to \cite{Ostrom2000},  has come to
be viewed as one of the most common representations of collective action problems
among other social dilemmas. Two recent contributions \cite{MerSanSzy2012,MerSanSzy2013} analyze boundedly rational agents interactions  in the $N$-person prisoner's dilemma using agent-based modeling to take into account of agents heterogeneity. 

Other social dilemmas have been considered; for example \cite{DalFornoMerloneWSC2013} obtains simulation results of interaction in the Braess Paradox using behaviors grounded on human participants behavior.

Note that a small number of locally interacting agents  seems  to be the most reasonable practical case often  contrary to many (theoretical) opinions.  In \cite {caram2010dynamic}   the agents  are supposed to interact  on a given market, with some ``strength''  depending on their size,   - a peer-to-peer competition. Their evolution is governed by a set of Lotka-Volterra dynamical equations, as for prey-predator problems. The fitness of the companies thus evolve according to the  value of their neighbors (in a continuum space, but through binary interactions only).  The role of initial conditions knowledge  is emphasized  in  \cite {caram2010dynamic}  following some analytical studying with a few agents (= companies).  Several behaviors emerge, going from one extreme in which a few agents compete
with each other, passing through oscillations,  reaching some   clustering state, up to the case of a ``winner takes the top'' state, and all others drop out. The  clustering phenomenon so obtained, in market language,  represents the natural segmentation into big, medium, and small ``players''. It  can happen that the segmentation can be extreme, even of the binary type. From a socioeconomic point of view, this means that a monopolistic situation is sometimes likely. Many examples come immediately in mind, but  are left for the reader to think over.

A line of research has used interactions among heterogeneous agents to model asset pricing and financial time series
in markets \cite{cerqueti2012role,cerqueti2010memory,cerqueti2008dynamics}.

\subsection{Interactions on Geometrical Structures} 
\label{subsec:GeomInt}
 When considering geometrical structures the line represents the simplest geometrical structure.
 Probably, in the bi-dimensional case the most common interaction using a geometrical structure take place on a grid. The most known example is given by the game of Life \cite{conway1970game,conway1982life}. Actually, the game of Life is more properly a cellular automaton, that is a dynamic discrete system that can be defined as a lattice
(or array) of discrete variables or ``cells'' that can exist in different states, \cite{Iltanen2012}.   The reader may refer to Chapter 4 in \cite{NorthMacal2007} for details on how the story of agent-based modeling  is related to  John Conway's Game of Life and Schelling's  housing segregation model \cite{Schelling1969}.

\begin{figure}[b]
\includegraphics[scale=.45]{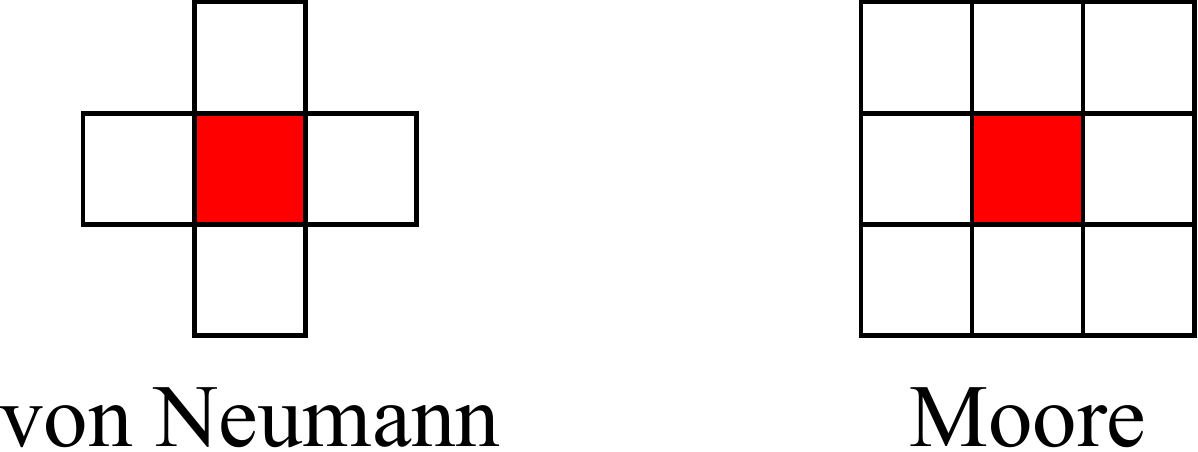}
%
%
\caption{Neighborhoods on bidimensional grids: von Neumann and Moore.}
\label{fig:2}       
\end{figure}

 On bi-dimensional grids several neighborhoods can be considered, the most common are the von  Neumann and Moore which are illustrated in Figure \ref{fig:2}. 
 
\begin{figure}[b]
\includegraphics[scale=.45]{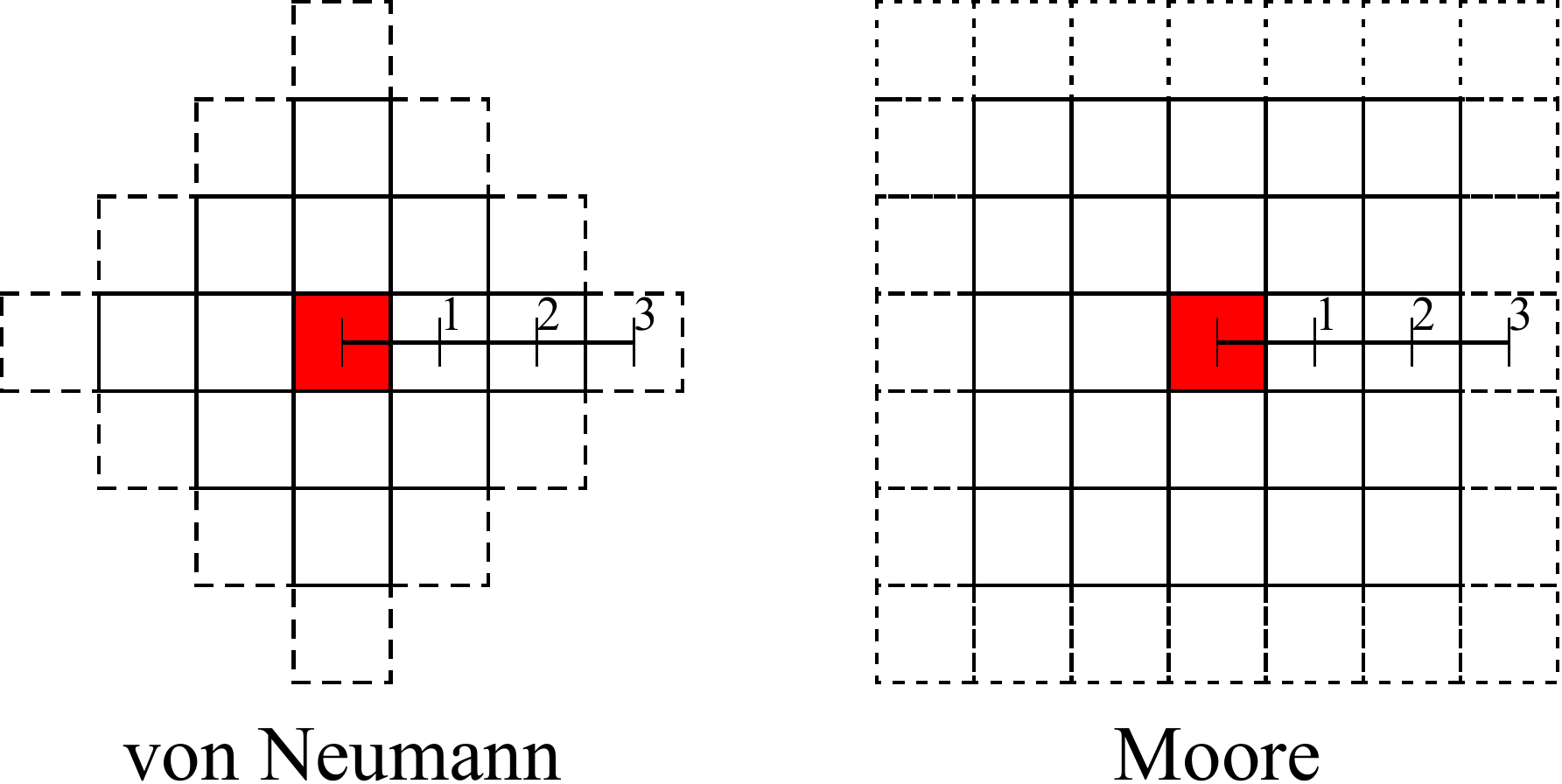}
%
%
\caption{von Neumann and Moore $d$-neighborhood with $d=1,2,3$.}
\label{fig:3}       
\end{figure}

 In this case interaction takes place only between adjacent cells and they can be called also von Neumann $1$-neighborhood and Moore $1$-neighborhood. It is possible to consider $d$-neighborhoods as illustrated in Figure\ref{fig:3}.

\begin{figure}[b]
\includegraphics[scale=.55]{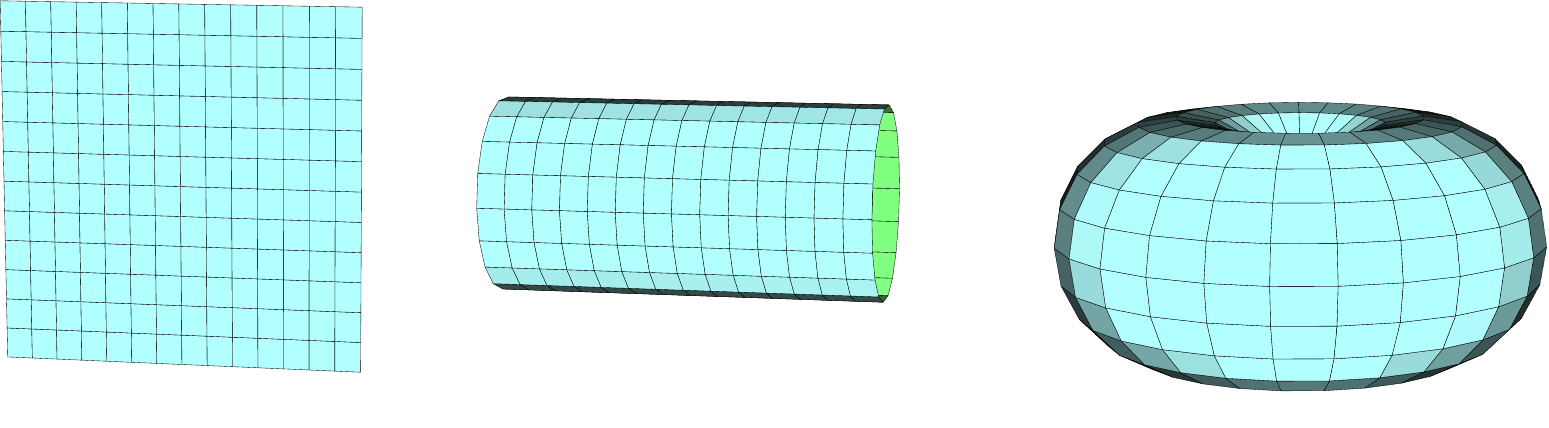}
%
%
\caption{Toroidal surface obtained when connecting the boundaries of a grid}
\label{fig:4}       
\end{figure}

 When the boundaries are connected a toroidal surface is obtained as illustrated in Figure \ref{fig:4}. In this case when  either the Moore or von Neumann neighborhood has a large enough radius and the likelihood of interaction does not depend on the agents distance the geometric structure is no longer important. In this case the agents can be thought as being part of a unique population as mentioned in Subsection  \ref{subsec:NoStr}.

 Being one of the simplest spatial interaction the one taking place on rectangular grids has been used in various applications, for a review in sociophysics see \cite{Stauffer2003}.  For example, interactions among players in two-persons game on a toroidal surface have been considered in    \cite{CerrutiGiacobiniMerlone2005}. Furthermore, in \cite{MerSziSzy2007}  finite neighborhood games have been considered with finitely many agents and with binary choices.  Among two-persons games a central role is played by the Prisoner's Dilemma. For example, \cite{NowakMay92} place agents on a two dimensional spatial array and observe the evolution of cooperation with deterministic players. Several interactions in organizations have the form of Prisoner's Dilemmas, therefore the modeling of such interactions has been used to explore cooperation in organizations as in \cite{DalFornoMerlone}; this analysis has been extended to consider personnel turnover in \cite{DalFornoMerloneNLDPLS} and in \cite{DalFornoMerloneCUBO} where a reward mechanism devised
to increase competition is introduced.

 In  \cite{DalFornoMerlone2006} the emergence of leaders is analyzed when considering interactions on a grid and distance models influence of potential leaders.

Finally, interactions on grids have been used to analyze the dynamics of industrial districts in \cite{MerloneTerna2006,MerloneEtAl2008}. 
 
In \cite{ausloos2003simple,ausloos2004evolution,ausloos2004model,ausloos2004reactive}, 
a few modern questions on economic policy: delocalization, globalization, cycles, etc., have been raised and tackled along modern statistical physics (Monte-Carlo) simulations on a rather simple but with realistic ingredient model. A highly simplified agent-based model has been first   introduced in  \cite{ausloos2004evolution}  and later developed in \cite{ausloos2003simple,ausloos2004model,ausloos2004reactive} providing a stylized geographical type of framework in order to touch upon answers on such fundamental socio-economic questions. Quantitative results similar to qualitative features are   found, {\it a posteriori} implying that  the model  ingredients are close to some accepted  common knowledge or stylized  reality.

The model is based on agents which interact with each other under various conditions, - several of them have been modified in the course of the investigations, thereby leading to several publications   \cite{ausloos2003simple,ausloos2004evolution,ausloos2004model,ausloos2004reactive}.

These contributions  are related to each other, but with different emphases, mainly depending on the evolution parameters. The lattice size(s), lattice symmetry (or symmetries), initial concentration(s), economic field time sequence(s), selection pressure, diffusion process rule(s), enterprise-enterprise ``interaction''(s), business plan(s), number of regions,  enterprise evolution law(s),  and economy policy time delay implementation are all presupposed to be known for the Monte-Carlo simulation.
  It is found that the model even in its simplest forms can lead to a large variety of situations, including: stationary solutions and cycles, but also chaotic behavior.

The model basically consists of 
\begin{enumerate}
\item  some{ \bf space} -- a square symmetry lattice, - sometimes divided into three ($k=$  $I$, $II$, $III$) {\bf equal size} regions, connected along a horizontal axis
\item  several {\bf companies}, initially randomly placed on  the   lattice sites, in an
\item  { \bf environment}   \index{environment} characterized by a real value,  so called {\bf external  (exogenous), field} $F_k \in [0,1]$,
\item  and a {\bf endogenous (internal) selection pressure} $ sel $;
\index{selection pressure}
\item each company ($ i $) is characterized by one real parameter $f_i \in [0,1]$,
 so called its {\bf fitness}.
\index{fitness}
\end{enumerate} 
The following set of actions is allowed to companies:
\begin{enumerate}
\item companies  evolve according to their  {\bf  survival probability }
 
$p_i = \exp (-sel\; |f_i -F_k | )$
 
compared at each time step to a predefined threshold $\theta$, which can be time and space dependent, though most of the time has been kept constant in time and within a given region
 \item companies {\bf may move} on the lattice 
one step at a time, horizontally or vertically,  thus 
in their von Neuman neighborhood; 
\index{von Neuman neighborhood}
\item if companies meet on a site, they may
\begin{enumerate}
\item either merge with a probability $b$,  and remain  on one of the two sites which was  occupied, with a new fitness value to start  a ``new life'' with,
\item or create a new company with the probability $1-b$,   creating  a $spin-out$ company, on some   (available) site\footnote{one such site is usually  available and  can be chosen arbitrarily}.
\end{enumerate} 
\end{enumerate} 
\index{mean field approximation}
This model is  called ACP and may be described in a mean field approximation (there is no spatial structure in such an approximation)
\cite{miskiewicz2004logistic,ausloos2004reactive} 
by introducing the
 distribution function of companies $ N(t,f) $,
which describes the number of companies having a given fitness $ f $ at time $ t $. The
system is then additionally characterized by the concentration of companies $ c(t) $.

The ACP model \cite{ausloos2003simple,ausloos2004model,ausloos2004reactive} looks like a reactive   lattice-gas (LG) system  \cite{simon1993statistical}; it  contains among its variants an adaptation of the Bak-Sneppen  (BS) model   \cite{bak1993punctuated}. The Bak-Sneppen model on a lattice has been used  to model the probability of occurrence of signals
commonly used in technical analysis \cite{rotundo2007microeconomic}. The same model, again applied on lattices and also  on scale-free
networks, has been used for the analysis of the firms'dynamics \cite{rotundo2009co}.

Note that the generation of new entities is more likely to occur 
in the case of a low concentration of companies than when this concentration is
high. The merging parameter describes the reversed dependency, i.e. merging
is more likely to occur in the case of a high density of companies than if the
density is low. 

Note also  that the three region  geographical-like  space was introduced in order to mimic north-south or east-west problems, allowing for a third  ``regional continent'' in the process, for the purpose of some generalization. In fact,  the model can be studied in one region only. In the simulations in fact, the companies are supposed to be only in region I at first, - in order to thermalize the system.   Thereafter,  the border between region I and region II  opening looks like the Berlin wall fall,
permitting the motion of firms into regions II, then III.  At the same time,   in some simulation, the external
field was allowed to change  in region I, assuming a new value $F_I$, different from those in region II
($F_{II}$) and III ($F_{III}$)\footnote{Usually the  field value was kept constant in the simulations, except for the mentioned drastic change. However  some generalization is of interest, looking for relaxation and memory effects.}.

A few fundamental results can be outlined. One   indicated that,
 depending on initial conditions, a cyclic process can be found, but  a decay of the number of companies in some region can occur.
  If companies are allowed to have some strategy,  based, e.g., on some information about the location, concentration or (/and) fitness of their neighbors, complex situations can occur. Interestingly it has been found that the ``reaction'' (or delay) time before implementing a decision (according to the strategy) induces a steady state, a cyclic state or a chaotic state.  This is both interesting and frightening, since the outcome depends on the real time scale and on initial conditions (usually poorly known).
  
One can also observe and measure a tendency  of the regional  concentrations toward some equilibrium state, governed by the external field and the threshold.  Interestingly, a bonus of  the ACP model is that it contains parameters which can be used in scaling empirical data, like the Monte-Carlo time between two field changes, - which are like changes in government policies.

An unexpected and interesting feature has been found:
the local changes of the environment is leading to sharp variations, almost discontinuous
ones, in the fitness and company concentrations, in particular when the field gradient is strong between regions.  A lack of self-organization is thus seen at region borders\footnote{The evolution of an economy, in which the functioning
of companies is interdependent and depends on external conditions, through natural
selection and somewhat random mutation is similar to bio-evolution}.  One can imagine its meaning if a largest set of regions is considered;

Note that the selection pressure looks like a temperature in thermodynamics   A critical value for order-disorder stares was found in    \cite{ausloos2004evolution}. This is somewhat relevant for estimating conditions for the creation of new enterprises. Note that most of the studies pertained to conditions on the ``best adapted'' companies; in view of the present crisis and  bankruptcy of many companies, the opposite case, could be investigated!

The ACP model has only been studied as a (Ising, spin 1/2) model. Yet, the LG approach allows already some increase in the number of  degrees of  freedom, characterizing the agents. However, it would be of interest to go beyond the (Ising, spin 1/2) model, and  still increase the number of degrees of freedom, using real company measures (income, stocks, benefits, ...) for pursuing the investigations.  Moreover the forces behind a strategy, like  bargaining power and/or market threats, labor and/or transaction costs could be usefully map into  endogenous fields, by raising the scalar field into a vector field, which could be space and time dependent as well. However the delay time  effect and the  initial conditions knowledge constraints might make the investigations rather looking like  conjectures.

Let it be observed that in contrast to the three region model presented in \cite{CommendatoreKubin2013},  the ACP model does not consider transportation costs as an ingredient;  in fact, this cost is considered to be constant in [24]. It would be of interest to investigate further this constraint.

\subsection{Interactions on Networks}
\label{subsec:NetwInt}
 More recently several ABM have been considering interaction on Networks.  
In \cite{AlamGeller2012}  the relationships between agent-based
social modeling and social network analysis are discussed throughly. 

It is important to consider that networks in ABM can play different roles. The physical space 
on which the interaction takes place can be modeled by a network and obviously all the structures described in the previous sections can be represented as networks. In this case agents are not represented by the network nodes, rather they interact using the network  as a physical support. For example, according to \cite{GilbertTroitzsch2005}, the model of segregation proposed by Schelling in \cite{Schelling1969} can be considered as a migration model, i.e., a cellular automata where actors are not confined to a particular cell. In this case, the line where interaction takes places can be viewed also as network.

 When agents are not identified with the nodes and move on physical space interacting depending on their position, a social interaction network is defined.

For example,  when all nodes are connected we have a a complete graph
 (see  \cite{WassermanFaust}), in this case  the network structure becomes trivial and we are back to the cases described in  Subsection  \ref{subsec:NoStr}.

 The relationship between interactions and the network can be thought at least in two directions. On one side it is possible to assume that the pattern of interactions define the connection among nodes. For example it can be assumed that two nodes are connected if and only if at least an interaction takes place; in this case  we have a dichotomous network. By contrast, when the strength of links is given by the number of interactions we have valued networks  (see  \cite{WassermanFaust}). 
 
  On the other hand it can be assumed that the connections of the Network nodes are given and that interactions can take place only between connected nodes.  In this case, the network defines the structure on which the interation takes place. Finally, it is important to observe that the geometrical structures described in Subsection \ref{subsec:GeomInt} can be modeled using networks. In this sense networks are a generalization of the geometrical structures so far considered.

When considering the first case, i.e., only some nodes are connected either directionally or nondirectionally it is possible to study particular interactions as those of a supervised team as in \cite{DalFornoMerloneWSC2012,DalFornoMerlone2003,DalFornoMerlone2007JSC,DalFornoMerloneIJIEM}. These contributions ground agent behaviors on human participants experiments. 
  
There are also  models in which both kinds of networks are considered. For example, \cite{DalFornoMerlone2007B,DalFornoMerlone2008} consider interactions between team members using two different networks. The first one is the knowledge network which  is necessary to work interaction between agents; the second network is the one which describes the work interactions which actually take place among agents. One of the interesting findings of this contribution is that starting with completely connected knowledge network, i.e., with no structure among the agents, does not allow the emergence of the more productive teams. On the other hand, a balanced expansion of the knowledge matrix is necessary for having agents working on the most productive projects. These finding have important consequences when considering social networks and the number of connections, which are related to the Dunbar's number \cite{Dunbar1992,Dunbar1993}. In fact, according to this Author there is a cognitive limit to the number of people with whom one can maintain stable social relationships. 
 
The network of team workers has been further examined in \cite{DalFornoMerloneQQ} where  structural balance  is introduced. In \cite{DalFornoMerlone2009} the network interaction is used to examine the role of social entrepreneurs in the emergence of cooperation. The role of social networks for the emergence of wage inequality is studied in \cite{dagem13} using an agent-based  analysis.

Interactions on different networks are considered in the literature, see \cite{stauffer2003sociophysics,duran2005evolutionary,axtell2001effects,krzakala2008potts,ochrombel2001simulation,sousa2005reshuffling,stauffer2012biased}; for a review of complex social networks the reader may refer to  \cite{Vega-Redondo2007}. According to some Authors none of the standard network models fit well with sociological observations of real social networks. An interaction model grounded in social exchange theory is  proposed in \cite{pujol2005} and  \cite{hamill2009} discuss how standard network models fit well with sociological observations of real social networks and proposes a new model to create a wide variety of artificial social worlds.

Even starting with an equal distribution of goods at the beginning of  a closed market, on a fixed network, with free flow of goods and money, it can be shown  that the market stabilizes in time  \cite{ausloos2007model}. This occurs faster    for small markets than large ones, and even  for systems in which money is steadily deduced from the market, e.g. through taxation during exchanges\footnote{ It was found that  many characteristic features are quite similar to the  ``no taxes'' case, but again the differences are mostly seen in the distribution of wealth,  -- the poor gets poorer and the rich gets more rich.}. It has also been  found that the price of goods decreases, when taxes are introduced, likely due to the less availability of money. In fact, in extreme situations, prices may not represent actual values, as is somewhat of common understanding.

Even though the model in \cite{ausloos2007model} is the most  simple money-goods exchange model, (there are only two ``parameters'':  the size of the market, or in other words, the number of agents,  and the initial amount of goods and money attributed to each agent),  the results are somewhat  indisputable, though frightening. Any complication of the rules, thereafter, is based on ``politics'', but no one knows if any change in rule produces a better situation.   But it can serve some of the agents instead of others....

 A more detailed discussion can be found in \cite{VarelaEtAl} in this volume.

\section{ABM for Modeling Geographical Distribution of Economic Activities}
\label{sec:ABMMod}

Understanding which factors determine the geographical distribution of economic activities and the differences in output, income, productivity or 
growth rates across regions is one of the most pressing issues in economics both from a theoretical and a policy perspective. Since these distributions are shaped 
by patterns of spatial interactions and factor flows that evolve over time, gaining a good understanding of the mechanisms responsible for the emergent outcomes asks for the
consideration of dynamic spatial processes and interactions. The main body of (theory-based) economic research in this area relies on models that capture the spatial geographical structure in a very simple form and do not provide an explicit representation of the underlying dynamic processes by relying on the assumptions the economy is always in equilibrium. Most prominent in this respect is the large body of literature on new economic geography (NEG), where standard models assume a simple Core-Periphery structure 
and dynamics is only considered in terms of the movement of short-run equilibrium outcomes towards a long-run equilibrium (see e.g. \cite{cometal06} and the survey on NEG in this volume). Although these approaches provide a large range of important insights into the mechanisms responsible for aggregation respectively disaggregation of economic activity they typically do not address issues like the dynamic stability of the projected long run outcome with respect to non-equilibrium adjustment processes of the economy, the effects of path dependencies in the adjustment dynamics, the effects of different kinds of spatial frictions on the goods and factor markets, the differences between short- and long run effects of policy measures on spatial distributions or implications of firm and household heterogeneities within and between regions. Agent-based spatial models allow to address such issues, but arguably the potential of an agent-based approach in this domain has not been fully exploited yet. Although several agent-based macroeconomic models with spatial structure have been developed in recent years (apart from the work discussed below, see e.g. \cite{wolfetal13}) and spatial agent-based models have been used for policy analysis in agricultural economics for some time (e.g. \cite{ber01,hapetal08,filetal09}), overall there is relatively little agent-based work dealing with the dynamic processes leading to spatial distributions of activities and spatial economic policy issues. This section reviews some of the existing work, where subsection \ref{subthe} focuses on literature exploring general agglomeration mechanisms and subsection \ref{subpol} covers studies dealing with spatial policy issues. Although no claim of completeness is made for the coverage of these subsections this review makes clear that there is substantial room for more research in economic geography using an agent-based approach. 
 
\subsection{An Agent-Based Perspective on New Economic Geography Models Out of Equilibrium \label{subthe}}
In a series of papers  Fowler has used agent-based simulations to address the question in how far the qualitative findings of standard New Economic Geography models can be transferred to a setting which does not assume equilibrium a priori. In \cite{fow07} he develops an agent-based model which sticks as closely as possible to the assumptions of the standard Core-Periphery model without assuming that all markets always clear and that all households and firms act optimally given current factor costs and prices (e.g.~firms use mark-up pricing based on past wage costs). Furthermore, in contrast to the standard NEG approach, where the number of firms in a region is determined by the size of the local labor market, in \cite{fow07} firm exit-entry and relocation processes are explicitly modeled. The simulation results reported in \cite{fow07} show that the agent-based version of the Core-Periphery model in a large fraction of runs generates full agglomeration of workers in the long run, even for parameter constellations where the analytical version would predict relatively equal distribution of activities across regions. Furthermore, for parameter settings where also the analytical model yields full agglomeration the agent-based model in the majority of runs ends up with full agglomerations in regions different from the one predicted by the standard NEG analysis. Fowler identifies the interplay of worker and firm relocation dynamics as the source for these discrepancies. Whereas workers have incentives to move to regions where wages are highest, for firms regions are most attractive where factor costs, in particular wages, are low. This leads to rationing of firms on the labor market, regional unemployment and overall path dependencies yielding different agglomeration locations in different runs and in most cases to outcomes that are not compatible with the results of the equilibrium analysis. The conclusion in \cite{fow07} is that it remains quite unclear by which dynamic processes that are viable also out of equilibrium the equilibrium points considered in the Core-Periphery literature can be attained. 

In \cite{fow11} this line of work is extended by introducing adjustment processes that allow to smooth the discrepancies between labor supply and demand in a region which arise due to the relocation dynamics described above. In particular, it is assumed that firms do not change regions but adjust employment due to hiring and firing. In addition firms might exit the market as conditions warrant and new firms might enter into a region where the gap between supply and demand is particularly high. It is shown that in cases where all workers are identical this version of the model to a large extend reproduces the results of the equilibrium analysis. However, the picture changes significantly if heterogeneity among firms and workers is introduced. In particular, in a scenario where reservation prices of workers are heterogeneous the agent-based model converges in less than 40\% to the prediction of the standard NEG Core-Periphery analysis. As pointed out in \cite{fow11} the scenarios with heterogeneous agents are however exactly the cases one might expect to see in the real world. Hence these findings suggest that substantial additional work is needed to gain a better understanding of the dynamic process generated by the forces underlying the NEG literature. Agent-based models seem a natural tool to undertake such work. 

Whereas the contributions by Fowler rely on agent-based models whose structure closely resembles that used in the NEG literature, a few other authors have studied the spatial dynamics of factor flows and economic activities based on stylized models focusing on particular types of economic activities and particular mechanisms. In \cite{ottetal01} a grid is considered on which firms and households search for locations. The focus is on the implications of the interplay of heterogeneous types of agents characterized by different decision rules governing their location search. Also the impact of changes of the radius that individuals take into account in their search is considered. An agent-base model which focuses on the interplay of learning through social interaction, creativity and location decision of workers is developed in \cite{spe12}.  
It is analyzed how specialization between regions and the agglomeration of creative activity is influenced by different aspects like the educational system or migration incentives. In \cite{yanett12} the emergence of spatial patterns of economic activity is analyzed from the perspective of firms which grow at different rates, might spawn spin-outs and relocate. The interplay of Marshall and Jacobs externalities with congestion effects is captured in an agent-based multi-level, multi-scale model. It is demonstrated how different preferences with respect to spatial proximity lead to different spatial agglomeration patterns. 
Finally, in the urban dynamics literature agent-based models have been developed to study agglomeration patterns on an aggregate level without capturing economic transactions on the micro-economic scale. For example, the SIMPOP model provides a rather disaggregated representation of spatial
interactions by reconstructing in a multi-agent simulation model the trade flows
between a large number of urban regions that are characterized by their economic portfolio (see \cite{breetal10}). 
 

The use of ABM for land use has become quite popular as \cite{MatthewsEtAl2007} illustrates.
 There are several reviews on these applications see  for example 
see a recent review in \cite{ParkerEtAl2002}. In this section we will be more interested in considering the spatial interactions in terms of economic localization. Other models consider ABM for dynamic disaster environment management, see \cite{Fiedrich2004}.

\subsection{Labor Flows, Regional Growth and Effects of Convergence Policy: Insights from Agent-Based Analyses \label{subpol}}
The empirical observation that regional differences in economic activity and per capita income are not only in many instances
persistent over time, but may even grow due to different regional growth rates is not only of great interest for economists, but is also a major concern
for policy makers. Cohesion policies aiming at the reduction of regional inequalities are among the most intensively funded policy areas in the European Union\footnote{In the period from 2007 to 2013, 347 bn Euros have been spent for cohesion policies, which makes about 36\% of the total EU budged.} and also numerous individual countries run programs to foster the catch-up of economically lagging regions. Also, there has for a long time been a vivid political debate about the implications of regional differences with respect to institutional setups (e.g. the organization and flexibility of the labor market) and trans-regional factor flows, in particular labor and foreign direct investment, on regional growth and convergence dynamics. In spite of this large empirical importance of cohesion policy issues, systematic analyses of the short-, medium-
and long-run effects of concrete policies from a theoretical perspective are largely missing. One of the reason for this lack of model-based policy analysis in this area might be that standard dynamic equilibrium models typically are too abstract to capture important characteristics of the policy measures under consideration. Also, a focus on long run steady states and balanced growth rates does not allow to study the short and medium run implications of the policies as well as potentially arising path dependencies, which seem to play an important role in empirical explanations of persistent regional differences in economic performance and policy effects. Finally, a comprehensive understanding of the effects of policy induced changes in factors like local skill endowments, labor market flexibility, labor mobility, infrastructure or support for technological development of firms must take into account the feedback between the direct effects of such measures and their implications for spatial factor flows, induced changes of firm behavior inside and outside the region as well as changes in the (distribution of) characteristics of agents (e.g. skills) in the different regions. Capturing the dynamics of these feedback requires an explicitly dynamic spatial model which includes these different sector and their connections as well as firm behavior.  

\subsubsection{The Eurace@Unibi Model}
In a series of papers the multi-regional agent-based Eurace@Unibi model has been used to address policy questions of the kind discussed above in a dynamic spatial context. 
The Eurace@Unibi model is based on the agent-based
macroeconomic simulation platform developed within the EU-funded EURACE project.\footnote{This
project (EU IST FP6 STREP grant 035086) was carried out by a
consortium lead by S. Cincotti (University of Genova), H. Dawid (University of
Bielefeld), C. Deissenberg (Universit\'e de la Mediterran\'ee), K. Erkan
(TUBITAK National Research Institute of Electronics and Cryptology), M.
Gallegati (Universit\`a Politecnica delle Marche), M. Holcombe (University of
Sheffield), M. Marchesi (Universit\`a di Cagliari), C. Greenough (STFC -
Rutherford Appleton Laboratory).} After the completion of the EURACE project in
2009 a number of authors have extended and altered the model
substantially in numerous directions leading to the current version denoted as the Eurace@Unibi model. Extensive discussions of the Eurace@Unibi model can be found in \cite{daetaldocu12} or in \cite{daetaloup}. In these contributions it is also shown that the Eurace@Unibi model is able to reproduce a large number of empirical stylized facts on different levels of aggregation.  

\begin{figure}[p]
\includegraphics[width=16cm]{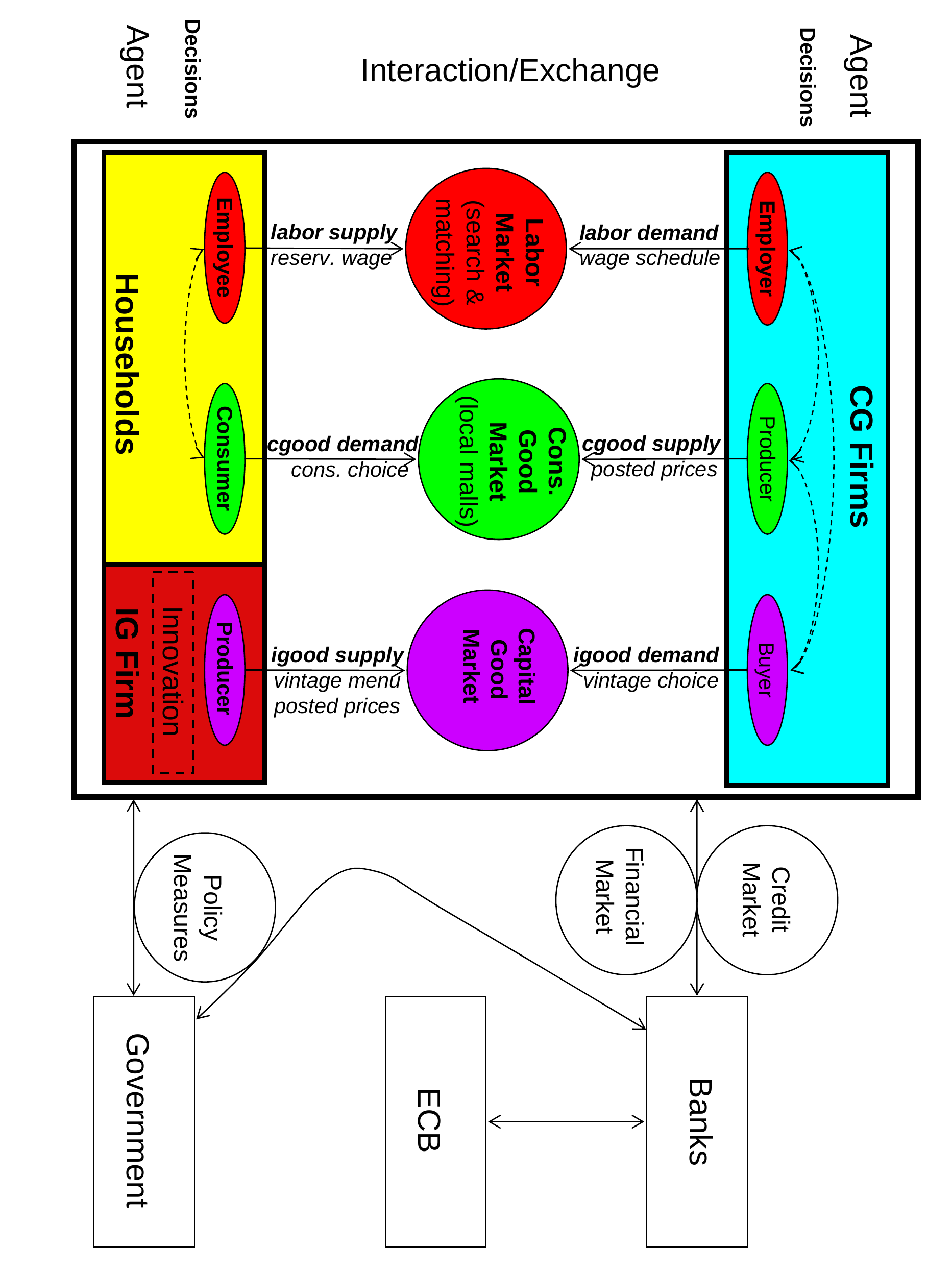}
\caption{Overview over the Eurace@Unibi model} \label{fig:over}
\end{figure}

The model describes an economy containing labor, consumption goods ('cgoods'), capital
goods (abbreviated as 'igoods' for investment goods), financial and credit markets in a regional context. The economy is
inhabited by numerous instances of different types of agents: firms
(consumption goods producers and capital goods producers), households and
banks. Each of these agents is located in one of the regions.
Additionally, there is a single central bank and a government that collects
taxes and finances social benefits as well as potentially some economic policy
measures, where policies might differ between regions. Finally, there is a
statistical office (Eurostat) that collects data from all individual agents in
the economy and generates aggregate indicators according to standard procedures.
These indicators are distributed to the agents in the economy (which might use
them e.g. as input for their decision rules) and also stored in order to
facilitate the analysis of the simulation results. A graphical overview over the crucial parts of
the model is given in figure \ref{fig:over}.

Capital goods of different quality are provided by capital goods producers
with infinite supply. The technological frontier (i.e.~the quality of the best
currently available capital good) improves over time, where technological
change is driven by a stochastic (innovation) process. Firms in the
consumption goods sector use capital goods combined with labor input to
produce consumption goods. The labor market is populated with workers that
have a finite number of general skill levels and acquire specific skills on
the job, which they need to fully exploit the technological advantages of the
capital employed in the production process. Every time when consumption goods
producers invest in new capital goods they decide which quality of capital
goods to select, thereby determining the speed by which new technologies
spread in the economy. Consumption goods are sold at local market platforms
(called malls), where firms store and offer their products and consumers come
to buy goods at posted prices. Labor market interaction is described by a
simple multi-round search-and-matching procedure where firms post vacancies,
searching workers apply, firms make offers and workers accept/reject. Wages of
workers are determined, on the one hand, by the expectation the employer has
at the time of hiring about the level of specific skills of the worker, and,
on the other hand, by a base wage variable, which is influenced by the (past)
tightness of the labor market and determines the overall level of wages paid
by a particular employer. Banks collect deposits from households and firms and
give credits to firms. The interest that firms have to pay on the amount of
their loan depends on the financial situation of the firm, and the amount of
the loan might be restricted by the bank's liquidity and risk exposure. There
is a financial market where shares of a single asset are traded, namely an index
bond containing all firms in the economy. The dividend paid by each share at a
certain point in time is determined by the sum of the dividends currently paid
by all firms. This simple representation of a financial market is not suitable
to describe speculative bubbles in the financial market, but captures
important feedbacks between firm profits and households income, in a sense
that fluctuations of dividends affect only the income of a particular subgroup
of households, namely the owners of shares of the index bonds. The central
bank provides standing facilities for the banks at a given base rate, pays
interest on banks' overnight deposits and might provide fiat money to the government.

Firms that are not able to pay the financial commitments declare illiquidity.
Furthermore, if at the end of the production cycle the firm has negative net worth, the firm is insolvent and insolvency bankruptcy is declared. In both cases it goes out of business, stops all productive activities and all employees loose their jobs. The firm writes off a fraction of its debt with all banks with which it has a loan and stays idle for a certain period before it becomes active again.

The spatial extensions of the markets differ. The capital goods market is
global meaning that firms in all regions buy from the same global capital good
producer and therefore have access to the same technologies. On the
consumption goods market demand is determined locally in the sense that all
consumers buy at the local mall located in their region, but supply is global
because every firm might sell its products in all regional markets of the
economy. Labor markets are characterized by spatial frictions determined by
commuting costs that arise if workers accept jobs outside their own region. It
is assumed that firms have access to all banks in the economy
and, therefore, credit markets operate globally.

In contrast to dynamic equilibrium models, where it is assumed that the
behavior of all actors is determined by maximization of the own
(inter-temporal) objective function using correct expectations about the
behavior of the other actors, agent-based simulation models need to provide
explicit constructive rules that describe how different agents build expectations and take their different decisions based on the available
information, which typically does not include information about the exact structure of their economic environment. Actually, the need to provide such rules is not only based on the
basic conviction underlying these models, that in most economic settings
actual behavior of decision makers is far from inter-temporally optimal
behavior under rational expectations, but also on the fact that in most models
that incorporate heterogeneity among agents and explicit interaction protocols
(e.g.~market rules) the characterization of dynamic equilibria is outside the
scope of analytical and numerical analysis. The choice of the decision rules in the Eurace@Unibi model is based on a
systematic attempt to incorporate rules that resemble empirically observable
behavior documented in the relevant literature. Concerning households, this
means that for example empirically identified saving rules are used and
purchasing choices are described using models from the Marketing literature
with strong empirical support. With respect to firm behavior the
'Management Science Approach' is followed, which aims at implementing relatively simple
decision rules that match standard procedures of real world firms as described
in the corresponding management literature. 
A more extensive discussion of the Management Science approach can be found in \cite{daha11}.

A first analysis based on the Eurace model which addresses the question how regional growth can be fostered is carried out in \cite{daetal08}. This paper contributes to the debate whether activities to strengthen technological change should be centered on stronger regions, weaker regions, or better be uniformly distributed. The concrete policy measure under consideration is an increase in the level of general skills of workers in a region. Due to fact that higher general skills induce faster acquisition of specific skills of workers and the observation that firms can only fully exploit the quality of their physical capital stock if their workforce has
appropriate specific skills (which is captured in the Eurace@Unibi model), such a policy measure should have an impact on the technology choices and the productivity of firms in a region, thereby inlfuencing regional growth.

Under the assumption that the flow of workers between regions is hindered by substantial spatial frictions (which might be due to commuting costs or legal restrictions) the simulation experiments show that the concentration of policy measures in one region in the short run triggers stronger overall growth in the
economy compared to a uniform allocation of policy measures across both regions, but that such spatial concentration of the policy effort has relatively detrimental effects on long run growth.

 The findings are driven by the relatively low mobility of labor compared to that of consumption goods, which in the long run leads to an incomplete substitution of production in the low skill (less supported) region with the production in the high skill (more supported) region. We refer to \cite{daetal08} for a detailed discussion of the economic mechanisms underlying these results.  
  
Subsequent work in \cite{daetal09} shows that the assumption of substantial spatial frictions made in \cite{daetal08} is indeed crucial for the qualitative findings obtained in that paper. In \cite{daetal09} it is demonstrated that under the empirically hardly relevant assumption of zero commuting costs (i.e. workers are completely indifferent between working in their own or some other region as long as the same wage is offered), no significant differences between the effects of the policy types emerge. If the frictions in labor mobility are positive but small the spatially concentrated policy induces faster long-run growth than the uniform one. With a spatially concentrated policy a self-reinforcing cycle of capital and labor investments, emerges, which is triggered in the region where the policy is concentrated. The origin of this cycle is an initial asymmetry in labor costs and prices induced by the combination of a geographically concentrated skill-upgrading policy and (small) spatial labor market frictions. 

The focus in \cite{daetal12} is on the question how different policies of opening up labor markets accompanying an integration process of goods markets affect output and consumption dynamics in regions that start(ed) from different levels of economic development.
It is explored to which extent spatial frictions with respect to labor mobility may have positive or detrimental effects on overall and region-specific variables related to the well-being of their citizens in the medium and long run. 

The simulation experiments using the Eurace@Unibi model show that total output in the whole economy is lowest for closed regional labor markets. All policies that mimic an opening up of labor markets result in higher total long run output, where the differences
between these policies in terms of long-run output is negligible. 
Effects on regional output however differ between all four policies. In particular, convergence between the regions is strongest if no labor flows between the regions are allowed. This scenario corresponds however to relatively low total output. Among the policies inducing higher total output the one generating the highest labor flows between regions reduces the inequality between regions the most.   

Another concrete European regional policy issue is analyzed in \cite{daetal13} using the Eurace@Unibi model. The focus of the paper is on the effectiveness of different types of cohesion policies with respect to convergence of regions. 
Motivated by the main instruments used by the European Union (European Fund for Regional Development, European Social Fund)
the effects of two types of policies are compared: technology policy, providing subsidies for firms in an economically lagging region who invest in technologies at the technological frontier, and human capital policy, inducing an improvement of the distribution of general skills in the workforce in the target region. Two different setups are considered where in the first setup the labor markets
are fully integrated such that there are small frictions and all workers have almost unhindered access to both local labor markets. In the other setup the labor markets are completely separated and workers can only work in their
home region.  

The main results of the analysis are that the human capital policy is only
effective in terms of fostering cohesion if labor markets are separated. If labor
markets are integrated, output actually falls in the lagging region at which the
policy is targeted. Technology policies speed up convergence for integrated
and separated labor markets.
The negative implications of the human capital policy under open labor markets arise although the direct goal of improving the level of specific skills and of the vintage choice in the lagging region is reached. The negative effects of the policy for region two are due to the induced changes in the labor market tightness in that region, which have implications for wage dynamics, (relative) goods prices, demand shifts and investments.  

The different analyses of spatial policy issues discussed in this subsection use the possibilities opened by the use of an agent-based 
simulation approach to capture how different kind of spatial frictions and spatial flows of goods and production factors affect
regional economic dynamics. They capture these effects in the presence of heterogeneous firms and workers, which implies that agents
are differently affected by the spatial flows and that the characteristics of the spatial flows (e.g. the skill distribution of commuting workers) depends on the distribution of agents within the regions and across regions. The discussed results are first contributions in this direction but highlight the potential of the use of agent-based models for an improved understanding of policy effects in a spatial setting.       

\section{Conclusion}
The interest on the potentiality of ABM approach to model complex interactions is evident from the number and the quality of recent publications. This approach seem to be an important tool to model spatial inequalities evolution  through time as it can take into account both the complex patterns determined by economic, geographical, institutional and social factors and the non-linearities in the decision processes of the agents. In particular parsimoniously detailed  model of the interaction between the relevant actors will provide the policy makers some important tools to simulate the consequences of their decisions. Since the the New Economic Geography approach,  describes economic systems as very simplified spatial structures, in this chapter we provided a classification and an analysis of the spatial interaction between agents. Determining the kind of spatial interaction will be the first step to build a model to approach the uneven geographical distribution of economic activities.

\end{document}